\documentclass[sigconf]{acmart}
\usepackage{hyperref}
\usepackage{hyperref} 
\usepackage{graphicx}
\graphicspath{{./images}}
\usepackage{float}
\usepackage{xcolor}
\usepackage{ragged2e}
\usepackage{siunitx}
\usepackage{indentfirst}
\AtBeginDocument{%
  \providecommand\BibTeX{{%
    \normalfont B\kern-0.5em{\scshape i\kern-0.25em b}\kern-0.8em\TeX}}}
\setcopyright{acmcopyright}
\copyrightyear{2020}
\acmYear{2020}
\acmDOI{10.1145/3417990.3420208}

\copyrightyear{2020}
\acmYear{2020}
\setcopyright{acmcopyright}\acmConference[MODELS '20 Companion]{ACM/IEEE 23rd International Conference on Model Driven Engineering Languages and Systems}{October 18--23, 2020}{Virtual Event, Canada}
\acmBooktitle{ACM/IEEE 23rd International Conference on Model Driven Engineering Languages and Systems (MODELS '20 Companion), October 18--23, 2020, Virtual Event, Canada}
\acmPrice{15.00}
\acmDOI{10.1145/3417990.3420208}
\acmISBN{978-1-4503-8135-2/20/10}
\renewcommand\footnotetextcopyrightpermission[1]{} 
\begin{document}

\title{Low-code Engineering for Internet of things: A state of research}

\author{Felicien Ihirwe*, Davide Di Ruscio**, Silvia Mazzini*, Pierluigi Pierini*, Alfonso Pierantonio**}
\email{name.surname@intecs.it*, (davide.diruscio & alfonso.pierantonio)@univaq.it**}
\affiliation{%
  \institution{ITS Lab (Intecs Solutions Spa)*, University of L'Aquila**}
  \state{Italy}
}
\renewcommand{\shortauthors}{Felicien et al}

\begin{abstract}
\textbf{Developing Internet of Things (IoT) systems has to cope with several 
challenges mainly because of the heterogeneity of the involved sub-systems and 
components. With the aim of conceiving languages and tools supporting the 
development of IoT systems, this paper presents the results of the study, which 
has been conducted to understand the current state of the art of existing  
platforms, and in particular low-code ones, for developing IoT systems. By 
analyzing sixteen platforms, a corresponding set of features has been 
identified to represent the functionalities and the services that each analyzed 
platform can support. We also identify the limitations of already existing 
approaches and discuss possible ways to improve and address them in the future.}

\end{abstract}

\maketitle

\section{Introduction}\label{sec:introduction}
According to European Union CONNECT Advisory Forum report \cite{CAF}, IoT 
promises to be one of the most disruptive technological revolutions since the 
advent of the Internet as projections indicate that more than 50 billion humans 
and objects will be connected to the Internet by the end of 2020. As we 
experience in daily life, now we see more and more intelligent traffic lights, 
advanced parking technologies, smart homes, and intelligent cargo movement. 
This is due to the rising adoption of artificial intelligence (AI), and 5G 
infrastructure is helping the global IoT market register an increased growth.

In 2018, the IoT market was valued at \$190 billion and it is anticipated to 
reach \$1,102.6 billion  by 2026. The IT industry is rapidly adopting 
cloud-based 
solutions, which are contributing to the growth of the market. 
Even though we see the 
spike in IoT technologies with much complexities and a huge amount of 
processing power, there are still even more sophisticated challenges lying 
behind the implementations of such systems. To highlight a few, IoT is facing a 
huge scale of heterogeneity in terms of devices, users, data sources, and 
communication modes that need to be combined to form a fully sensed 
application. This leads to the issue of standards as new technologies come in 
and change the game.  

The current trend of low-code development platforms (LCDP) catches people's 
attention due to their capabilities in easing the development of fully 
functional applications mostly cloud-based. The primary goal of LCDP is to 
facilitate people with less or no experience in software engineering to develop 
business applications using simple graphical user interfaces 
\cite{Sanchis_2019}. Lowcomote project \cite{lowcomote} aims to push that 
advancement to a more technical and sophisticated era of "Low-code Engineering" 
by employing the concrete basic engineering principles to the modeling world. 
This will be done by the merge of model-driven engineering, cloud computing, 
and machine learning techniques.

In this paper we want to show the current state of research on model driven 
engineering approaches for IoT by taking into account low-code development 
platforms in particular. We present the results and the findings that 
have been done by analyzing sixteen IoT development platforms. They are 
divided into two categories considering their basic implementation mechanisms. 
In particular, the first category consist of tools based on the Eclipse technologies such as Eclipse Modeling 
Framework(EMF), Graphical Modeling Framework(GMF), and Papyrus environment. The second category is a collection of tailor-made platform referred as low-code development platforms. 

The study has been performed in three main steps: first, 
we conceived a taxonomy consisting of features characterizing the 
studied IoT development platforms. Then, such features are used to evaluate the functionalities and the services supported by each analyzed platform. As a last step we identified some weakness of the analyzed platforms to pave the way 
toward a low-code platform for developing IoT systems.  We have also identified 
some limitations of already existing approaches and discuss possible ways to 
improve and address them in the future. 

The rest of this paper is organized as follows:  Section \ref{sec:ioteng} makes 
an overview the IoT engineering core concepts. Section \ref{sec:state} presents 
the currently available languages and tools supporting the development of IoT 
systems. Section \ref{sec:taxo} presents a taxonomy we have 
defined to support the comparison of the considered IoT development approaches. 
By relying on the outcomes of the performed comparison, Section 
\ref{sec:discussion} discusses the strengths and limitations of existing 
approach in order to identify future research directions. Section 
\ref{sec:conclusion} concludes the paper and mentions some future work.

\section{Engineering IoT systems} \label{sec:ioteng}
In the past, IoT referred to the advent of bar-codes and Radio Frequency 
Identification (RFID), helping to automate inventory, tracking, and to support 
basic identification needs. The current wave of IoT is characterized by a 
strong verve 
for connecting sensors, objects, devices, data and applications 
\cite{iotrefence}. A fully IoT system is generally complex and its development 
typically requires many 
players of various expertise and several stakeholders with different 
responsibilities. IoT is regarded as a collection of automated procedures 
and data, integrated with heterogeneous entities (hardware, software, and 
personnel) that interact with each other and with their environment to reach 
common goals \cite{bruno}. 

\begin{figure}[t]
    \centering
    \includegraphics[width=8.6cm]{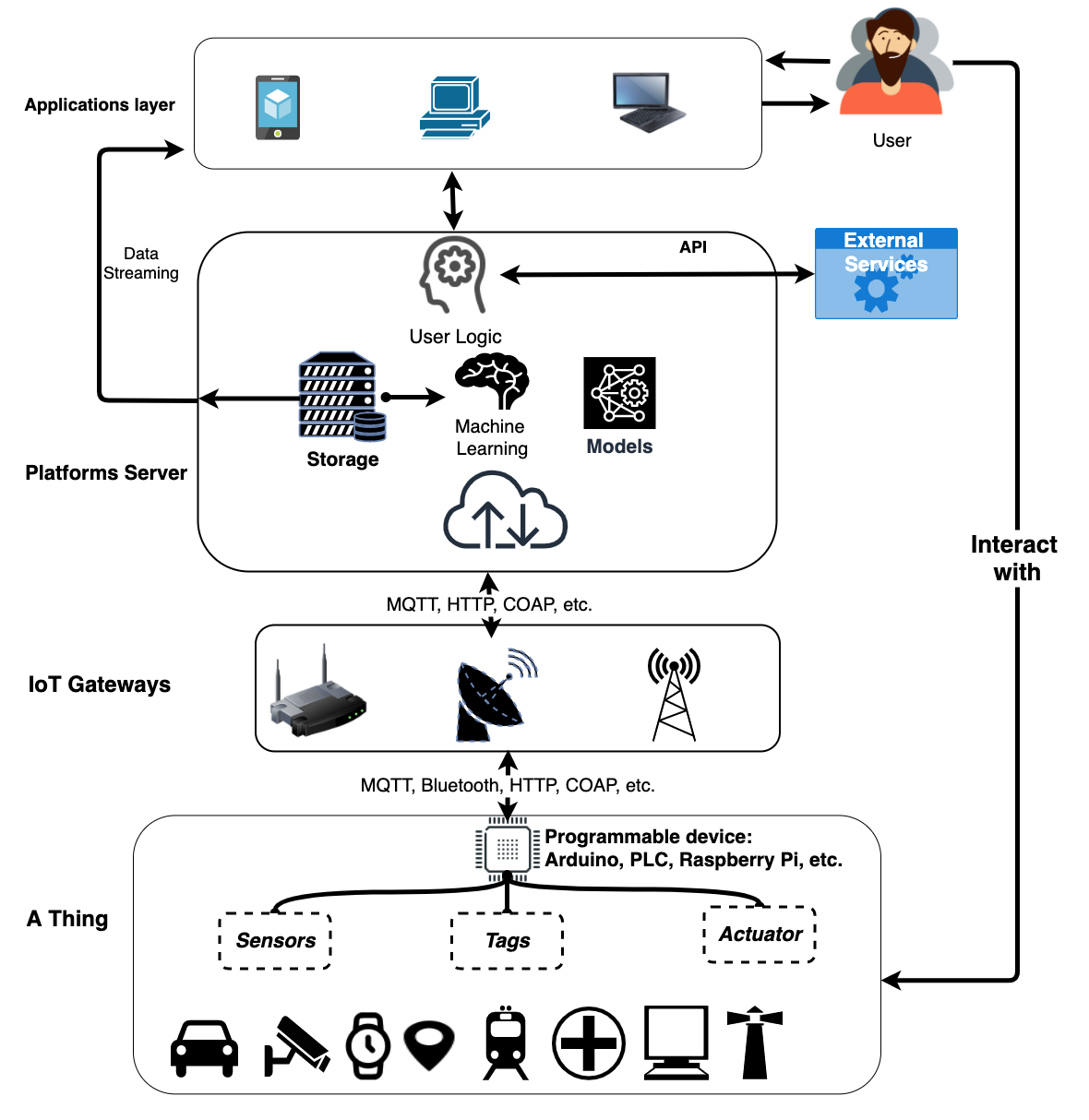}
    \caption{IoT system building blocks}
    \label{fig:IoTArchitecture}
\end{figure}

Figure \ref{fig:IoTArchitecture} shows a high-level architecture of a typical 
IoT system. \textit{A thing} is a combination of on-board devices including 
\textit{sensors}, \textit{tags}, \textit{actuators}, and \textit{physical 
entities} like cars, watches, etc. Data is generated from a sensor or a tag 
attached to the physical entity the user is interested in. A 
computing device (such as an Arduino, a Pycom, a Raspberry Pi, etc.) collects data and 
send them to the nearby gateway using some well-known protocols such as Z-Wave, 
MQTT, HTTP, Bluetooth, Wi-Fi, Zigbee, etc. 

The \textit{Gateway} component acts as a bridge between the physical and 
digital worlds.  Note that in some cases devices and gateways can make some 
simple computation logic and respond to some events without the need for 
further processing. The platform server is a combination of processing and 
storage resources on the cloud. At this stage, data can be streamed, analyzed, 
or manipulated for meaningful information to be communicated back to actuators, 
users, or third parties services.  

%
%
To support the development of complex IoT systems, several standards and tools 
have been proposed over the last years \cite{sysml4iot1}. Standards like 
ISO/IEC/IEEE 15288\footnote{\url{https://www.iso.org/standard/63711.html}} 
have been in use to evaluate the quality, efficiency, life-cycle of different 
approaches. In  \cite{IotrefAnalysis} an IoT reference model (ITU-T 
Y.2060)\cite{iotrefITU} is proposed by an International Telecommunication Union 
(ITU) and presented with respect to other four reference architectures 
developed in the context of the IoT-A \cite{iotrefence}, WSO2 
\cite{wso}, Korean RA \cite{iotrefITU} and Chinese 
\cite{chinaRefModel} projects. 


\begin{figure}[t!]
    \centering
    \includegraphics[width=8.6cm]{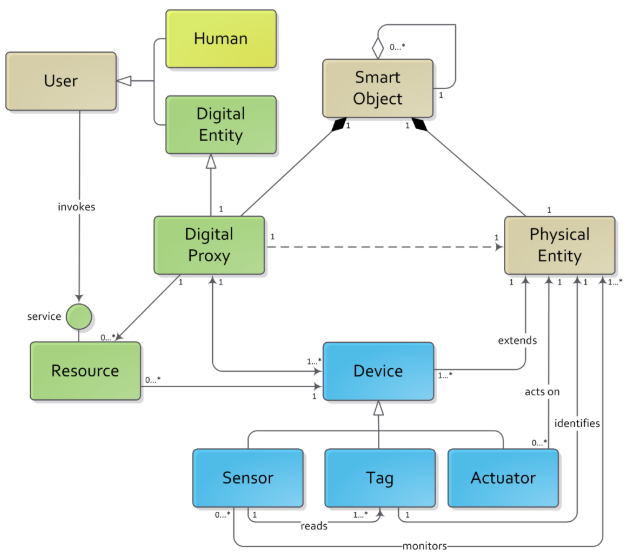}
    \caption{IoT conceptual metamodel \cite{serbanati2011building}}
    \label{fig:iotmetamodel}
\end{figure}

Figure \ref{fig:iotmetamodel} shows a conceptual representation of the elements 
shown in Figure \ref{fig:IoTArchitecture}. The physical properties of the 
associated 
\textit{Physical entities} are captured through \textit{Sensors}, whereas the 
modification of physical properties of associated \textit{Physical Entity} is 
performed through the use of \textit{Actuators}. \textit{Physical Entity} can 
be represented in the digital world by a \textit{Digital Entity} which is in 
turn a \textit{Digital Proxy}.  \textit{Digital Proxy} has one and only one 
\textit{ID} that identifies the represented object. The association between the 
\textit{Digital Proxy} and the \textit{Physical Entity} must be established 
automatically. A \textit{Smart Object} has the extension of a \textit{Physical 
Entity} with its associated \textit{Digital Proxy} which then talk to the user 
through by providing or requesting resources. The external services are invoked 
by the user which can be human or third-party software.

\section{IoT development approaches} \label{sec:state}
This section makes an overview of languages and tools to develop IoT systems by distinguishing those that implement modeling principles from proper low-code development platforms. The purpose of this study is to highlight the current state of the art in model based development for IoT systems. To ensure the authenticity of our search, we have conducted the search following the three main steps as listed below.

First, we conducted a manual search on google scholar\footnote{\url{https://scholar.google.com/}} using keywords such as \textit{{“Model driven engineering IoT"}}, \textit{{"MDE IoT"}}, \textit{{"Model driven engineering Internet of Things"}}, and \textit{{"MDE Internet of Things”}}.  

Second, we manually analysed and selected the papers following the content in the abstracts and conclusion.s Note that we were more interested in tools that can sell on the market not just approaches with some minor implementation. To this end, we have selected approaches satisfying the following criteria:
\begin{enumerate}
    \item Availability of supporting tools;
    \item More recent than 2010;
    \item Use of MDE as underlying technique;
    \item Referred  by at least three scientific publications.
\end{enumerate}

Third, we conducted a paper mapping phase starting on the approaches found in the previous step. We mostly considered what approach a paper refers to or what refers to it. This was not an easy task. We went manually through each of the selected approaches by consulting through several publisher sites such as IEEE, ACM and Springer. This manual search was basically conducted using the tool's name as the main searching keyword.

On the lowcode development side, we mainly considered the LCDPs selected in a recent Gartner report \cite{gartner} that focuses on promising low-code platforms. We have also added a few LCDPs obtained by conducting a manual search on google using \textit{{“Lowcode Development platform IoT"}}, and \textit{{“Lowcode Development platform Internet of Things”}} keywords.
The "Lowcode development platform" is a more recent term, the results found were mostly present in \cite{gartner}. The additional LCDPs was selected due their popularity and the fact that they are capable of implementing the basic IoT elements depicted in Fig. 
\ref{fig:IoTArchitecture}.

\subsection{MDE approaches}

This section presents an overview of modeling languages and supporting tools that authors have identified by analyzing existing literature in the MDE research field.

\textbf{MDE4IoT framework \cite{mde4iot}:} It is a tool to support the 
architectural modeling and self-adaptation of emergent configurations regarded 
as Things. This approach was validated by modeling the Smart Street Lights use 
case \cite{mde4iot}. MDE4IoT is a combination of different UML Domain Specific 
Languages as profiles and validated code generators. MDE4IoT uses a combination 
of different viewpoints to capture the system functionalities from hardware to 
software which help to accurately perform model-to-model transformations. The 
approach relies on the UML-MARTE profile for modeling hardware components as 
well as allocations of software to hardware. Run-time adaptations are meant to 
be performed automatically by specific in-place model transformations.

\smallskip
\textbf{SysML4IoT \cite{sysml4iot1}:} It is a SysML profile based on the IoT-A 
reference model presented in \cite{iotrefence}. The \textit{SysML2NuSMV 
translator} is also available to generate NuSMV\footnote{\url{http://nusmv.fbk.eu/}} specifications out 
of input SysML models with the aim of supporting the verification of 
Quality of Service (QoS) properties by taking into account environmental 
constraints. SysML4IoT has been adopted in \cite{bruno} in the context of the 
\textit{IDeA–IoT DevProcess} \& 
\textit{AppFramework}. The IoT DevProcess has been extended from OOSEM\footnote{\url{https://www.incose.org/incose-member-resources/working-groups/transformational/object-oriented-se-method}} 
standards and tailored to support IoT application design. To 
support the activities of IoT DevProcess through SysML4IoT, all hardware 
devices and software components are precisely identified in the system model as 
stereotypes. The same tool has been adopted in \cite{mahmud} to develop IoT 
self-adaptive systems endorsing the public/subscribe paradigm to model the 
communication with other systems.

\smallskip
\textbf{Papyrus for IoT \cite{papyrus4iot}:} It is a modeling tool based on 
Papyrus. The approach has been already used to develop a smart IoT-based home 
automation system as a benchmark application to showcase the capability of the 
solution. Authors suggest the extension of Papyrus Moka \footnote{\url{https://wiki.eclipse.org/Papyrus/UserGuide/ModelExecution}}  to 
perform model simulations. Concerning the deployment of the modeled systems, 
authors make use of Prismtech's Vortex as a dynamic platform to discover and 
deploy microservices, and MicroEJ as the target operating system. 

\smallskip
\textbf{UML4IoT \cite{uml4iot3}:} It is an MDE based tool developed for 
industrial automation systems (IASs).  A model-to-model transformation has been 
developed to automatically transform the mechatronic components to Industrial 
Automation Things (IAT). UML4IoT has been exploited to automate the 
generation process of the IoTwrapper, a software layer that is required on top 
of the IEC61131\footnote{\url{https://plcopen.org/iec-61131-1}} cyber-physical 
component to expose its functionality to the digital IoT ecosystem. The authors 
developed a LWM2M application protocol running on top of the CoAP communication 
protocol. 

\smallskip
\textbf{IoTML \cite{brainiot}:} It is a tool developed in the context of the 
BRAIN-IoT project\footnote{\url{http://www.brain-iot.eu/}}. BRAIN-IoT presents 
an integrated modeling environment to ease the rapid prototyping of smart 
cooperative IoT systems based on shared models. The BRAIN-IoT architecture 
mainly consists of three macro-blocks: the BRAIN-IoT modeling Framework, the 
Marketplace, and the federation of BRAIN-IoT Fabrics. The BRAIN-IoT modeling 
environment includes the IoTML modeling tool, which is implemented as a Papyrus 
profile. The IoTML tool permits to specify models that can be uploaded to the BRAIN-IoT marketplace to foster their future reuse. 

\smallskip
\textbf{IoTLink \cite{iotlink}:}  It is a development toolkit based on a 
model-driven approach to allow inexperienced developers to compose mashup 
applications through a graphical domain-specific language. Modeled applications 
can be easily configured and wired together to create an IoT application. 

To support interoperability with other services, authors implemented custom 
components like ArduinoSerial for Arduino connectivity, SOAPInupt, RESTInput, 
MQTTInput, etc. At runtime, the tool generates connections by using the 
Drools\footnote{\url{https://www.drools.org/}} engine to poll the rules from a 
database repository, which allows developers to deploy and change deployment 
rules at runtime.

\smallskip
\textbf{FRASAD \cite{frasad}:}  It is a model driven framework to develop IoT 
applications. FRASAD relies on a node-centric software architecture and a 
rule-based programming model. FRASAD has been developed atop of Eclipse 
EMF/GMF and consists of a graphical modeling language, a code generator and 
other supporting tools to help developers design, implement, 
optimize, and test the developed IoT applications. 

\smallskip
\textbf{ThingML \cite{MDEjungle}:} It consists of a modeling language and 
supporting tools already employed to develop commercial e-health solutions 
\cite{MDEjungle}. The ThingML language 
combines well proven software-modeling constructs aligned with UML 
(statecharts and components) and an imperative 
platform-independent action language to construct the intended IoT 
applications. The tools include an advanced multi-platform code generation 
framework that supports multiple target programming languages such as C, C++, 
Java, Arduino and JavaScript. The tool is open to the community as an Eclipse 
plugin. 

\smallskip
\textbf{ThingML+ \cite{thingml+}:} It is an extended version of the ThingML 
tool\cite{MDEjungle} designed for IoT/cyber-physical systems with the aim of 
addressing machine learning needs. The authors presented a new approach of 
using machine learning techniques to tackle the issue of IoT communications and 
behavioral modeling which was normally being done using state machines. The 
techniques have been integrated at the modeling level 
and code generation level. The approach has been developed in the context of 
the  
ML-Quadrat\footnote{\url{https://www.quadrat.ac.uk/quadrat-projects/}} research 
project.

\smallskip
\textbf{AutoIoT \cite{AutoIoT}:}  It is a framework that allows users to model 
their IoT systems using a simple JSON file. The process starts by modeling the 
system using the graphical interface generated from GMF. When the modeling 
phase is completed, AutoIoT loads the content of the model as a JSON file to be 
validated and transformed into Python objects using the 
Pydantic\footnote{\url{https://pydantic-docs.helpmanual.io/}} library. After 
that, the framework finally delivers these objects to an appropriate 
\textit{Builder} that performs model-to-text transformations to generate a 
ready-to-use IoT server side application. The Prototype Builder generates a 
Flask application written in Python, HTML, CSS, and Javascript. The generated 
server-side application communicates with IoT devices and third-party systems 
through MQTT, Rest API, and WebSockets.

\smallskip
\textbf{IoTSuite \cite{IoTSuite}:} It is a tool suite consisting of the 
following components: \textit{i)} an \textit{editor} to support the application 
design phase by allowing stakeholders to specify high-level descriptions of the 
system under development; \textit{ii)} an 
ANTLR\footnote{\url{https://www.antlr.org/}} 
based \textit{compiler} that parses the high-level specification and generate 
an IoT framework; \textit{iii)} a \textit{deployment module}, which is 
supported by mapper and linker modules;  \textit{iv)} a \textit{runtime 
system}, which leverages existing middleware platforms and it is responsible 
for the distributed execution of the modeled IoT application. The current 
implementation of IoTSuite targets both Android and JavaSE-enabled devices and 
makes use of an MQTT-based middleware.

\smallskip
\textbf{CAPS \cite{caps1}:} It is an architecture-driven modeling tool 
based on Eclipse EMF/GMF designed to enable architects, system engineers, and 
cyber-physical space designers to capture the software architecture, hardware 
configuration, and physical space into views for a situational-aware CPS. To 
link together the modeled views, the authors introduced two 
auxiliary languages, denoted as Mapping Modeling Language (MAPML) and 
Deployment Modeling Language (DEPML). The authors used the Atlas Model Weaver 
(AMW)\footnote{\url{https://projects.eclipse.org/projects/modeling.gmt.amw}} to 
define relations among models and to create semantic links among model 
elements. The tool also has a \textit{CAPSml} code generation framework built 
on top of the CAPS modeling framework which transforms the CAPS model into 
ThingML\cite{MDEjungle} code which can then be used by the ThingML framework to 
generate fully operational code.

\smallskip
Authors in \cite{dslsmartcity} present a domain-specific language that can be 
used to model the structural and behavioral views of a smart city application. 
A metamodel was developed using EMF and the graphical interface of the DSL was 
generated using GMF. In their work, the authors used OCL (Object Constraint 
Language) to define constraints and uniqueness of elements in a model by 
introducing element rules. The approach fosters also the adoption of a 
repository of already modeled elements by taking care of device definitions and 
configurations.

\smallskip
In \cite{nodecentric} authors present a DSL designed to specify all aspects 
of a sensor node application, especially for data processing tasks such as 
sampling, aggregation, and forwarding. The proposed DSL offers a set of 
declarative sentences to express the behaviour of sensor nodes application such 
as \textit{sampling}, \textit{aggregating}, and \textit{forwarding} which is 
necessary for developing data-centric Wireless Sensor Network (WSN) 
applications. The tool is based on Eclipse GMF for specifying PIMs. The 
transformation from the PIM to 
nesC\footnote{\url{http://nescc.sourceforge.net/}} PSM models has been 
implemented by using the ATL tansformation 
language\footnote{\url{https://www.eclipse.org/atl/}}. Acceleo-based model-to-text transformations have been developed to generate final nesC source code of 
the modeled system.

\subsection{Low-code development platforms for IoT}

Low-code development platforms (LCDPs) aim at bridging the gap between 
experienced software developers and people with less or 
no experience in software engineering. LCDPs have shown their strengths in the 
development of software systems in four main market segments such as database 
applications, mobile applications, process applications, and 
request-handling applications. According to \cite{lowcomote}, IoT will be the 
next market segment for LCDPs. In this section, we make an overview of 
existing  LCDPs for IoT that take into account MDE concepts in their core implementations. 

\smallskip
\textbf{Node-RED}\footnote{https://nodered.org/}  is a programming tool 
specifically conceived in the IoT context, with the aim of wiring and 
connecting together hardware devices, APIs, and online services \cite{NodeRed}. 
It provides a cloud-based editor that makes it easy to connect together flows 
using the wide range of nodes in the palette that can be deployed to its 
runtime easily. Node-RED provides a rich text editor built on top of Node.js 
taking full advantage of its event-driven, non-blocking model. Node-RED can be run locally or on the cloud. Node-RED is platform agnostic and compatible with several devices  such as Raspberry Pi, BeagleBone Black, Arduino, Android-based devices. Node-RED 
also supports its integration with cloud-based resources such as IBM Cloud, 
SenseTecnic FRED, Amazon Web Services, and Microsoft Azure.

\smallskip
\textbf{AtmosphericIoT}\footnote{\url{https://atmosphereiot.com/}} provides 
IoT solution builders with languages and tools to build, connect, and manage 
embedded-to-cloud systems \cite{AtmosphericIoT}. Atmosphere IoT Studio offers a 
free drag-and-drop online IDE, to build all device firmware, mobile apps, and 
cloud dashboards. AtmosphericIoT connects devices from a range of wireless 
options including Wi-Fi, Bluetooth and BLE, Sigfox, LoRa, ZigBee, NFC, 
satellite, and cellular. This platform is entirely cloud-based but it offers 
downloadable artifacts.

\smallskip
\textbf{Simplifier}\footnote{\url{https://www.simplifier.io/en/}} is a low-code 
Platform for integrated business and IoT applications that enable user to 
create, manage, deploy, and maintain enterprise-grade 
SAPUI5\footnote{\url{https://www.guru99.com/sapui5-tutorial.html}} apps for 
web, mobile, and wearables. Simplifier uses a pre-built interface for 
bidirectional integration of existing SAP and non-SAP systems and leverage 
shop-floor integration with native IoT-interfaces (OPC-UA, MQTT) 
\cite{Simplifier}. It is provided as a web-based environment available 
on-premise or in the cloud. Simplifier permits to deploy applications on SAP 
Cloud Platform, SAP NetWeaver, as stand-alone, or on a dedicated Simplifier 
cloud.

\smallskip
\textbf{Mendix}\footnote{\url{https://www.mendix.com/}} is one of the popular 
low-code development platforms that offer significant enterprise 
characteristics especially attractive to large businesses\cite{gartner}. Its 
platform is  equipped for multi-cloud and hybrid computing, due to its 
support for on-premises, virtual private multi-cloud, and multitenant public 
cloud deployment options \cite{Mendix}. 

\smallskip
\textbf{Salesforce}\footnote{https://www.salesforce.com/} is a 
popular CRM low-code development platform that has been adopted through many 
different new technologies like AI, Machine Learning, and Cloud computing. 
Salesforce supports the rapid prototyping of IoT applications through the 
connection with the underlying Salesforce IoT cloud engines \cite{Salesforce}. 
This platform provides data visualization and event management through a visual 
set of rules and triggers on different data source components.

\section{Taxonomy} \label{sec:taxo}
In this section we introduce a taxonomy of terms, which can support the description and the comparison of different approaches for the development of IoT systems. By analyzing the languages and tools 
overviewed in the previous section, we identified and formalized their corresponding variabilities and commonalities in terms of a feature diagram.

These features were selected mainly based on the common understanding regarding the stages to be followed in the software development process, from requirement definition, system design, development, deployment and maintenance of a robust complex system. 
This was an iterative process where all authors were involved in the development of the proposed taxonomy. Findings related to the performed study are discussed in the next section.

\begin{figure}[b]
    \centering
    \includegraphics[width=8.6cm]{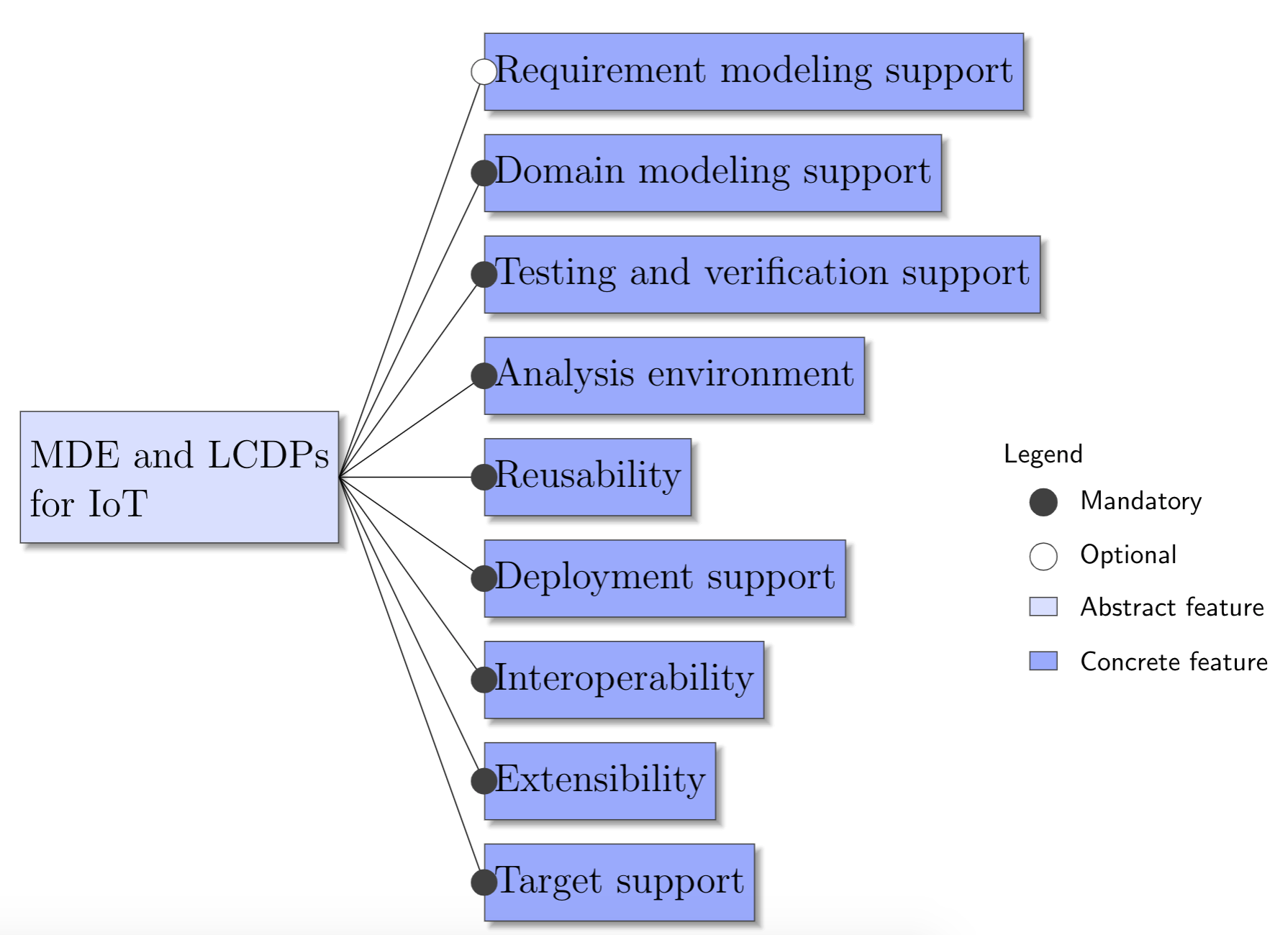}
    \caption{Feature diagram representing the top-level variation areas}
    \label{fig:feature}
\end{figure}

Figure \ref{fig:feature} shows the top-level feature diagram, where 
each sub-node represents a major point of variation. Table \ref{tab:findings} gives details about the taxonomy described in the following.

\smallskip
\textbf{\emph{Requirement modeling support:}} this group of features 
emphasizes the first stages of any MDE based development process. 
This evaluates whether a tool have an inbuilt requirement specification 
environment. Supporting this feature is very important because it helps 
keep track on whether the specified requirements are correctly 
implemented throughout the whole development. This also helps in 
requirements traceability and verification.

\smallskip
\textbf{\emph{Domain Modeling support:}} it refers to the kind of modeling tools the user is provided with e.g., if it is graphical or not, if it gives the possibility to model the static structure of system's blocks or components. Some of the analyzed systems provide modelers with behavior modeling capabilities to specify semantic concepts relating to how the system behaves and interacts with other entities (users or other systems). For instance, OMG based implementations of UML/SysML inherits all the modeling functionalities which include structural and behavioral diagrams. Additionally, we looked at if the tool can support system modeling through multiple view which in turn is considered as multi-view modeling support.

\smallskip
\textbf{\emph{Testing and verification support: }} It refers at whether a tool has inbuilt mechanisms to evaluate artifacts before deployment which 
can be done by conducting different verification checks. To be more specific this feature examines if the tool has a testing workbench, an inbuilt model checking and validation facility. This is very important as it ensures the system correctness and robustness which make system safe and secure. As IoT applications are in our everyday life, we are interested in developing systems that will causes no harm to users in case of a more and more sophisticated scenarios.

\smallskip
\textbf{\emph{Analysis environment:}} such a group of features is related to the capability of the considered environment to support different analysis checks for the intended system before its deployment. This can be done on different blocks or components of the 
system by checking on their responsiveness in case of failure, network loss, security breach, and so on. In this regard, we can feature dependability 
analysis, real-time analysis, and system quality of service in general.

\smallskip
\textbf{\emph{Reusability:}} this category illustrates whether the tool under analysis allows 
the export of artifacts for future reuse. This can be done on developed models or on generated artifacts. Reusability features are also related to the way artifacts are managed e.g., locally or by means of some cloud infrastructure.

\smallskip
\textbf{\emph{Deployment support:}} it is related to the ways developed systems are deployed and if in general generated artifacts are ready for deployment or not. To the best of our knowledge, this should be one of the important features to focus on when implementing a novel tool. We also looked at whether the development tool can be installed locally or on the cloud depending on client's interest. Finally, we looked on whether the tool or platform provides run-time adaptation mechanisms (on modeled or generated artifacts) to respond to contextual changes and thus, react accordingly.

\smallskip
\textbf{\emph{Interoperability:}} this feature examines the ability of a tool to exchange information either internally between components, expose or consume functionalities or information from external services e.g., by means of dedicated APIs. 

\smallskip
\textbf{\emph{Extensibility: }} the tool should provide the means for refining or extending the provided functionalities. In the case of 
modeling tools, such a feature is related to the  possibility of adding new modeling features 
and notations. 

\smallskip
\textbf{\emph{Target Support:}} It refers to the characteristics of the target infrastructure, which enables the execution of the modeled system. Sub-features are represented by the  \emph{underlying infrastructure} to exhibit the core 
technologies a tool relies on, \emph{target platform}, which presents different devices and 
platforms supported by the generated code, and \emph{code generation language} to refer the programming languages supported by the considered system.

\begin{table*}[ht]
    \centering
    \includegraphics[width=\textwidth]{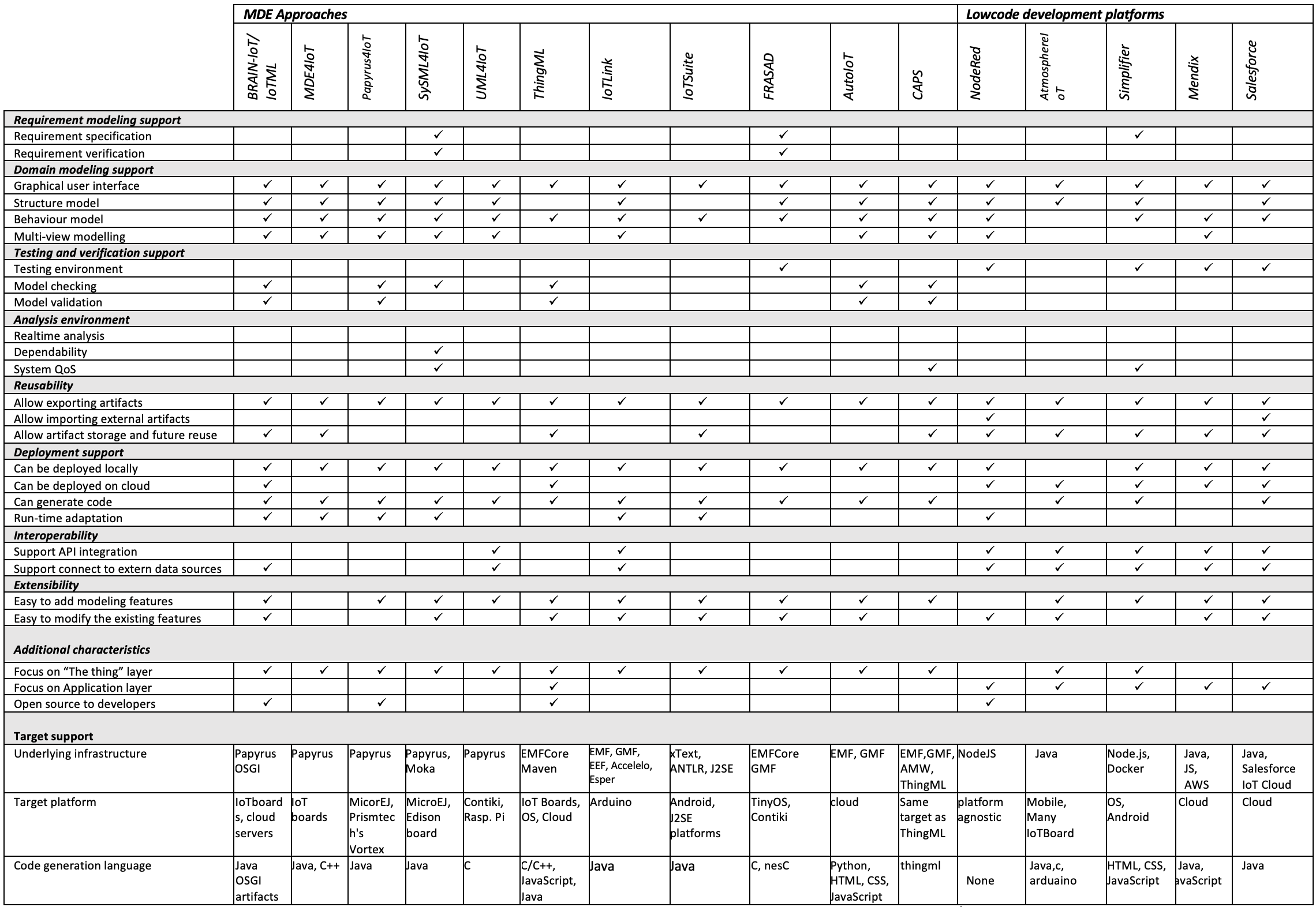}
    \caption{Taxonomy table}
    \label{tab:findings}
\end{table*}

Further than the previous features, we identified \textbf{\emph{"additional characteristics"}} that are orthogonal to the previously discussed elements. 
In particular, some  tools target early phases of development like system design, data acquisition, 
system analysis by focusing on the \textit{thing} behaviour. Some other tools target the application generation without taking much care of the data acquisition phases which can be done by integrating the developed system  with already implemented data source engines, etc. Another peculiar aspect is if the considered approach is available as open source or not having this an important impact on the possibility for the community to contribute to its development.

\section{Findings}\label{sec:discussion}
The elicited features, which have been discussed in the previous section, have been considered to  study and analyze 16 platforms. The selection has been done by considering all the platforms that cover the IoT 
modeling domain. 
According to Table \ref{tab:findings} most of the analyzed approaches rely on Eclipse and OSGi. We have realized a huge lack of focus on requirement specification except for SysML4IoT 
(as it extends SysML which enforces requirement specification) and FRASAD, which enforces the requirement specification at the PIM level using rules that can be tracked throughout. The huge lack of analysis support for almost all the tools selected is alarming. We think that it is highly important to analyze and verify the intended system's behavior before deployment as it gives developer indications of what may happen before deployment and help make any adjustment earlier enough. Moreover, we can see that most of the analyzed tools can be deployed locally especially concerning EMF-based tools but mostly all LCDPs are cloud-based with some of them being able to be run also locally. 
The main weakness that we discovered by the performed analysis are described below.

\smallskip
\noindent
\textbf{Lack of standards}: we noticed a lack of a standards to support the model-based development of IoT systems. We noticed that each tool proposes its own way of development by hampering interoperability possibilities among different platforms. This is due to the presence of many industrial players, which make the IoT meta-modeling convoluted. On the other hand, different research attempts proposed IoT reference models, which cover different development phases and 
perspectives. The IoT reference model presented in \cite{iotrefence} has been adopted by different tools 
\cite{sysml4iot1,papyrus4iot,iotlink} as a fundamental 
meta-model. This shows the potentials and benefits of having the availability of standards in such a complex domain. We believe that it as a good starting point, which needs to be further explored to better cover the interoperability dimension (e.g., to enable the possibility of interacting with third-party data resources in general).

\smallskip
\noindent
\textbf{Limited support of multi-view modeling:} we noticed that most of the approaches focus on single view modeling. In particular, except for CAPS, MED4IoT, Atmospheric IoT, and Mendix, the analyzed approaches use one specific view to model everything, which is not a good practice in general. Using multi-view modeling presents enormous benefits as it enforces separation of concerns: the system component is designed using a single model with dedicated consistent views, which are specialized projections of the system in specific dimensions of interest \cite{CHESSiot}. Multi-view modeling is regarded as a complicated matter to address for tailor-made low-code development platforms as they mostly focus on connecting dots aiming at having an application up and running. 

\smallskip
\noindent
\textbf{Limited support for cloud based model-driven engineering:} Moving model management operations to the cloud and supporting modeling activities via cloud infrastructures in general is still an open subject. From our study, we noticed that mostly low-code 
development approaches provide the option to run tools on cloud or on-premise. This is not yet the case of tools based on Eclipse EMF, which still requires local deployments. The research presented in \cite{MDEaaS} proposed a DSL as a Service (DSMaaS) as a 
solution to address the reusability of so many created DSL over the cloud. Other attempts like  MDEforge\footnote{\url{http://www.mdeforge.org/}} aim at realizing cloud based model manipulations \cite{mdeforge}. 

\smallskip
\noindent
\smallskip
\noindent
\textbf{Limited support for testing and analysis:} According to the performed study, very few tools care about the testing and analysis phases of the IoT system development process. There is still a big challenge regarding how to analyze IoT systems responsiveness before deployment. The complexity of the problem relies on the fact that IoT system involve human interaction, environment constraints and we have also to recognize the heterogeneity of the target platforms that makes it hard to depict the kind of analysis properties to address.

The above table and discussed findings should not be considered as a final reference but it should give a sense on the state of research in MDE based IoT development platforms. This is because we relied on published papers and official documented work which is considered as authentic but in some cases does not provide the full information on tool capabilities.

\section{Conclusion and Future Work} \label{sec:conclusion}
Model-driven engineering aim is to tackle different challenges faced in design, development, deployment, and run-time phases of software systems through abstraction and automation. In this study, we discussed state of the art on existing approaches supporting the development of IoT systems. In particular, we focused on languages and tools available in the MDE field and the emergent low-code development platforms covering the IoT domain. The analysis has been performed by conceiving a taxonomy, which has been formalized as a feature diagram presenting all the features of a typical modeling platform supporting the development of IoT systems. Different limitations have been identified in the analyzed platforms. As a future work, we want to continue the investigation on model driven engineering based IoT platforms by considering both the quantitative and qualitative aspects of the solutions developed following the two main development approaches presented in \ref{sec:state}. In the context of the Lowcomote project, we aim at addressing the presented limitations, especially by focusing on the early analysis possibilities that a low-code platform should provide to develop secure and trustworthy IoT systems.

\section{ACKNOWLEDGEMENT}
This work has received funding from the Lowcomote project under European Union’s Horizon 2020 research and innovation program under the Marie Skłodowska-Curie grant agreement n\si{\degree} 813884.

\bibliographystyle{ACM-Reference-Format}
\bibliography{bibliography}
\end{document}